\documentclass[pra,twocolumn,showpacs,amsmath,amssymb]{revtex4}

\usepackage[english]{babel}
\usepackage{graphicx}
\usepackage{dcolumn}
\usepackage{bm}

\usepackage[colorlinks=true]{hyperref} 
\usepackage{hyperref}
\hypersetup{
    hypertex,    
    pdftitle= {Quantum feedback by discrete quantum non-demolition measurements: towards on-demand generation of photon-number states},
    pdfsubject= {Article},
    pdfkeywords= { },
    pdfauthor= {\textcopyright\ Igor Dotsenko, 2009},
    pdfcreator= {\LaTeX\ with package \flqq hyperref\frqq},
    pdfpagemode=UseOutlines,
    hypertexnames=false,
    }
    \hypersetup{colorlinks=true}    
    \hypersetup{citecolor=blue}
    \hypersetup{linkcolor=blue}
    \hypersetup{bookmarks=true}
    \hypersetup{linktocpage=false}      
    \hypersetup{bookmarksnumbered=true} 
    \hypersetup{hyperfootnotes=false}

\newcommand{\Id}{\openone}
\newcommand{\EE}[1]{{\text{\large$\mathbb E$}}\left(#1\right)}
\newcommand{\bra}[1]{\left<#1\right|}
\newcommand{\ket}[1]{\left|#1\right>}
\newcommand{\tr}[1]{\text{Tr}\left(#1\right)}

\newcommand{\ngoal}{n_{\text{\tiny tag}}}
\newcommand{\rhogoal}{\rho_{\text{\tiny tag}}}

\newcommand{\nmax}{n_{\text{\tiny max}}}
\newcommand{\rhoreal}{\rho^{\text{\tiny real}}}
\newcommand{\rhohalf}{\rho_{k}}
\newcommand{\rhoaux}{\tilde{\rho}}

\newcommand{\Fconv}{F_{\text{conv}}}

\newcommand{\phiR}{\phi_{\text{R}}}
\newcommand{\Tcav}{T_{\text{cav}}}
\newcommand{\Tp}{T_{\text{a}}}
\newcommand{\nth}{n_{\text{th}}}
\newcommand{\etap}{\eta_{\text{a}}}
\newcommand{\etad}{\eta_{\text{d}}}
\newcommand{\etaf}{\eta_{\text{f}}}

\newcommand{\superoperDetected}[1]{\mathbf{F}_{\text{1} #1}}
\newcommand{\superoperUndetected}[1]{\mathbf{F}_{\text{2} #1}}
\newcommand{\superoperProjection}{\mathbf{P_{\text{}}}}
\newcommand{\superoperEvolution}{\mathbf{T_{\text{}}}}
\newcommand{\superoperInjection}{\mathbf{D_{\text{}}}}
\newcommand{\superoperNoprojection}{\mathbf{N_{\text{}}}}
\newcommand{\superoperMeasurement}{\mathbf{M_{\text{}}}}

\begin{document}

    \title{Quantum feedback by discrete quantum non-demolition measurements:\\ towards on-demand generation of photon-number states}
    \author{I. Dotsenko$^{1,2}$}
    \email{igor.dotsenko@lkb.ens.fr}
    \author{M. Mirrahimi$^3$}
    \author{M. Brune$^1$}
    \author{S. Haroche$^{1,2}$}
    \author{J.-M. Raimond$^1$}
    \author{P. Rouchon$^4$}

    \affiliation{
    $^1$Laboratoire Kastler Brossel, Ecole Normale Sup\'{e}rieure, CNRS, Universit\'{e} P.~et M.~Curie, 24 rue Lhomond, F-75231 Paris Cedex 05, France\\
    $^2$Coll\`{e}ge de France, 11 Place Marcelin Berthelot, F-75231 Paris Cedex 05, France\\
    $^3$INRIA Rocquencourt, Domaine de Vouceau, B.P.~105, 78153 Le Chesnay Cedex, France\\
    $^4$Mines ParisTech, Centre Automatique et Syst\`{e}mes, Math\'{e}matiques et Syst\`{e}mes,
    60~Bd.~Saint-Michel, 75272 Paris Cedex 06, France}

    \date{\today}

\begin{abstract}
We propose a quantum feedback scheme for the preparation and protection of photon number states of
light trapped in a high-$Q$ microwave cavity. A quantum non-demolition measurement of the cavity
field provides information on the photon number distribution. The feedback loop is closed by
injecting into the cavity a coherent pulse adjusted to increase the probability of the target
photon number. The efficiency and reliability of the closed-loop state stabilization is assessed by
quantum Monte-Carlo simulations. We show that, in realistic experimental conditions, Fock states
are efficiently  produced and protected against decoherence.
\end{abstract}

\pacs{42.50.Dv, 02.30.Yy, 42.50.Pq}


\maketitle

\section{Introduction}

It is now possible to realize ideal quantum measurements on individual quantum objects, for
instance atoms, ions and photons. Beyond being a tool to monitor the system's evolution, projective
quantum measurements can be used to prepare it in specific quantum states. For instance, an ideal
quantum non-demolition (QND) measurement of a field's photon number projects it onto a
photon-number (Fock) state \cite{Guerlin07}. However, due to the basic quantum indetermination of
the measurement outcome, measurement-induced state generation is not deterministic. Quantum
feedback control techniques \cite{Wiseman94} make it possible to overcome this limitation and to
produce quantum states on demand. These techniques generally combine weak quantum measurements with
a real-time correction of the system's state depending on the classical information extracted from
the measurements. Beyond preparation of specific states, these feedback schemes can also protect
them from decoherence, resulting from the coupling of the system with its environment.

In this paper, we propose a quantum feedback scheme for the on-demand preparation of Fock states
stored in a high-quality superconducting microwave cavity and for their protection against
decoherence. This scheme is designed to operate with an existing cavity-QED set-up
\cite{Deleglise08}. The feedback loop uses three steps. We first extract information on the photon
number distribution with a single circular Rydberg atom by a QND process. This atom interacts
dispersively with the cavity field and does not exchange energy with it, but experiences a light
shift proportional to the photon number. The measurement of this shift, with the help of a Ramsey
interferometer, provides information on the field intensity. It modifies the field state
accordingly through the quantum projection. This information is, in the second step, used to
estimate the new cavity field state through a quantum filtering process \cite{Belavkin92,Bouten07}.
In the third step, we correct the field state, using a coherent field pulse injected in the cavity.
The feedback law used to calculate the amplitude of this pulse is chosen by adapting to this
discrete situation the Lyapunov-based techniques proposed in Ref.~\cite{Mirrahimi07}. These three
steps represent the three basic components of any feedback loop: a sensor, a controller and an
actuator. In contrast to classical closed-loop systems, we use here a quantum sensor. However, both
controller and actuator remain classical. By iterating the feedback loop, we steer the cavity
towards any target Fock state. We also efficiently stabilize it against cavity decay.

Recently, a similar feedback scheme to generate Fock states of light in an optical cavity has been
proposed \cite{Geremia06}. It is based on a continuous monitoring of the mean number of photons.
Instead, we propose here to use a discrete QND measurement followed by a quantum filtering process
providing complete information on the field's density matrix. This precise knowledge of the field's
state gives us a much better insight into the feedback action. It could also be an asset in other
feedback schemes requiring, for instance, information on the field's phase in addition to its
intensity.

The paper is organized as follows. In Sec.~\ref{sec:setup}, we describe the elements of the set-up
and give realistic experimental parameters. We present the quantum-mechanical operators  describing
the evolution of the cavity state under measurement, decoherence and pulse injection.
Section~\ref{sec:ideal} is devoted to a detailed analysis of the quantum filter and of the feedback
law in an ideal situation. We first present the main elements of the feedback loop and describe the
tuning of the controller gain. We give qualitative arguments showing that the proposed strategy is
stable. The detailed mathematical proof of convergence and stability, relying on stochastic
Lyapunov techniques will be published elsewhere \cite{Mirrahimi09}. We finally present quantum
Monte Carlo simulations of closed-loop trajectories of the cavity field exhibiting the feedback
performances. In section~\ref{sec:real}, we take into account the known experimental imperfections
of the existing set-up. We modify accordingly the feedback algorithm and present extensive
simulations of its operation. We conclude in Sec.~\ref{sec:conclusion}.

\section{Experimental set-up}\label{sec:setup}

    \begin{figure}[t]
        \centerline{\includegraphics[width=0.48\textwidth]{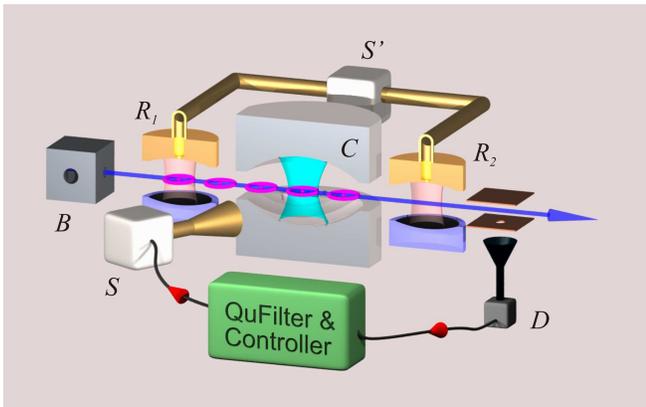}}
        \caption{Proposed quantum feedback scheme adapted to a microwave cavity QED set-up. $C$:~high-$Q$
        microwave cavity, $B$:~box producing Rydberg atoms, $R_1$ and $R_2$: low-$Q$ Ramsey cavities, $D$:
        atomic field-ionization detector, $S$ and $S'$: microwave sources coupled to $C$ and $R$'s, respectively.
        In a quantum filtering process, a
        real-time control system analyzes the results of QND measurements of the cavity field and computes the
        amplitude of a control injection pulse.}
        \label{fig:Setup}
    \end{figure}

Our quantum feedback algorithm is designed for the ENS microwave cavity QED
set-up~\cite{Raimond01,Deleglise08}. Its components are depicted in Fig.~\ref{fig:Setup}. The
microwave field to be controlled is confined in an ultra-high $Q$ superconducting cavity $C$
(damping time $\Tcav=0.13$~s). Rydberg atoms, flying one by one at 250 m/s across the cavity mode
and dispersively interacting with its field, are used as QND probes of light~\cite{Guerlin07}. They
are prepared in a superposition of two circular Rydberg states $\ket{e}$ and $\ket{g}$ (principal
quantum numbers 51 and 50, respectively) in the low-$Q$ cavity $R_1$, driven by the source $S'$, in
which they experience a resonant $\pi/2$-pulse. In the Bloch sphere representation, the spin
corresponding to the two-level atomic system then points along the $Ox$ direction in the equatorial
plane (states $\ket{e}$ and $\ket g$ correspond to north and south poles of the sphere,
respectively; we use a frame rotating at the atomic frequency).

The light shifts experienced by the non-resonant $\ket g\rightarrow\ket e$ atomic transition in the
cavity field result in a phase shift for the atomic state superposition. At the cavity exit, the
atomic spin points along a direction in the equatorial plane at an angle $\Phi(n)$ with the $Ox$
axis, correlated to the photon number $n$. The dephasing angle $\Phi(n)$ can be controlled by
adjusting the atom-cavity detuning $\delta$ and the interaction time. In the large atom-cavity
detuning regime, it is a linear function of $n$: $\Phi(n)\sim (\Omega_0^2/4\delta) n$, where
$\Omega_0/2\pi = 49$~kHz is the vacuum Rabi frequency. For the intermediate detuning values used in
the experiments, it is a more complex, albeit perfectly known growing function of $n$. A second
$\pi/2$ Ramsey pulse in $R_2$ (phase $\phiR$ with respect to that of the pulse in $R_1$) followed
by the atomic detection (in the $\{\ket e,\ket g\}$ basis) by the detector $D$ amounts to a
detection of the atomic spin along an axis at an angle $\phiR$ with $Ox$. It provides information
on the cavity field intensity.

Atoms are sent in the set-up at typical 250 $\mu$s time interval, much shorter than the cavity
damping time.  Note that such a macroscopic time interval is  well adapted to elaborate feedback
strategies since we have ample time to compute the state estimator and the feedback law between two
atomic detections. When no feedback action is performed, the information provided by a few tens of
atoms results in a measurement of the dephasing angle $\Phi(n)$ and, hence, in a projective QND
measurement of the photon number $n$ \cite{Guerlin07,Brune08}.

We are interested here instead in the ambiguous information provided by a single atomic detection.
Detection of the atomic state in $D$ projects the field, described initially by the density matrix
$\rho_\mathrm{}$, onto a new state $\rho_\mathrm{proj}$. Depending on the detected atomic state,
$\ket{s}=\ket{e}$ or $\ket{g}$, the back-action of the quantum measurement on the field is
described by
    \begin{equation} \label{eq:Proj}
        \rho_\mathrm{proj} =\frac{M_s\, \rho_\mathrm{}\, M_s^\dag}
                            {\tr{M_s\, \rho_\mathrm{}\, M_s^\dag}},
    \end{equation}
where the operators $M_s$  are given by
    \begin{subequations}
        \label{eq:MgMe}
        \begin{equation}
            M_g = \cos \left(\frac{\phiR + \Phi({N})}{2} \right),
        \end{equation}
    \vspace{-0.4cm}
        \begin{equation}
            M_e = \sin \left(\frac{\phiR + \Phi({N})}{2} \right).
        \end{equation}
    \end{subequations}
Here, ${N} = a^\dag a$ is the photon number operator with $a$ and $a^\dag$ the photon annihilation
and creation operators, respectively. The measurement operators $M_s$ are diagonal in a
photon-number basis (as is $N$). They thus preserve Fock states, illustrating the QND nature of the
measurement. The projected state \eqref{eq:Proj} is normalized by the probability $P_{s}=\tr{M_s
\rho M_s^\dag}$ of detecting the atom in state $\ket{s}$.

The cavity field can also be manipulated by injecting into the mode a coherent field pulse
generated by the resonant microwave source $S$. Its action is described by the displacement
operator $D(\alpha)=\exp(\alpha a^\dag - \alpha^* a)$, where $\alpha$ is the complex amplitude of
the injected field. The cavity field after displacement thus reads
    \begin{equation}\label{eq:injection}
        \rho_\mathrm{disp} = D(\alpha) \, \rho \, D(-\alpha).
    \end{equation}

A proper analysis of the feedback scheme requires to take into account all known imperfections of
the experimental set-up. Besides projective measurement \eqref{eq:Proj} and coherent evolution
\eqref{eq:injection}, the cavity state also evolves due to decoherence. The field is coupled to a
reservoir at non-zero temperature ($T=0.8$~K), and its dynamics is described by the master equation
\cite{Walls94}
    \begin{multline}\label{eq:L}
        \frac{d\rho}{dt} = \textbf{L} \rho \equiv
        - \frac{\kappa}{2}(1+\nth)(a^\dag a \rho + \rho a a^\dag - 2 a \rho a^\dag)\\
        - \frac{\kappa}{2}\nth(a a^\dag \rho + \rho a^\dag a - 2 a^\dag \rho a),
    \end{multline}
where $\kappa = 1/\Tcav$ is the cavity decay rate and $\nth = 0.05$ the equilibrium thermal
photon number.

Moreover, the circular Rydberg state preparation is a non-deterministic Poisson process. We perform
a pulsed excitation from a continuous thermal beam of Rubidium atoms, preparing atomic samples at a
$\Tp = 85\,\mu$s time interval. The mean number of circular atoms per sample is $\etap\approx 0.3$
(leading to an average time between atoms of 250 $\mu$s). The atomic time of flight between $C$ and
$D$ is 350~$\mu$s, corresponding to a delay of $d\!=\!4$ atomic samples. The probability for
detecting an atom present in a sample is  $\etad\approx 0.8$ due to finite detection efficiency in
$D$. Finally, non-ideal atomic state resolution of the field-ionization detector and finite
contrast of the Ramsey interferometer introduce a probability of erroneous state assignation of
$\etaf\approx 0.1$.

\section{Idealized experiment} \label{sec:ideal}

In order to describe the essential elements of the feedback scheme in a simple context, we first
consider in this Section an idealized experiment with no cavity decay ($\kappa = 0$), a
deterministic atomic preparation ($\etap=1$), no detection delay ($d=0$), and a perfect atomic
detection ($\etad=1$ and $\etaf=0$).

\subsection{Quantum filter} \label{ssec:ideal:filter}

The quantum filtering procedure \cite{Belavkin92,Bouten07} provides us with an estimate of the
field state by using all available information. Right after detection of atom number $k$ and before
injection of the corresponding control, it includes the initial state of the field $\rho_{0}$,
information obtained from all $k$ atoms detected so far (labelled $i=1\ldots k$) and all coherent
field injections performed in the former feedback loops (1 to $k\!-\!1$). Each of the elementary
processes (atomic detections and displacements) can be represented by super-operators acting on the
field's density matrix. Let us note $\superoperMeasurement_{i}$ that associated to the detection of
atom $i$ and  $\superoperInjection_{i}$ that corresponding to the displacement performed in the
$i^\mathrm{th}$ iteration of the feedback loop. Therefore, after detecting atom $k$, the state of
the field~is
    \begin{equation}\label{eq:rho_allKnowledge_ideal}
    \rho_{k} = \superoperMeasurement_k
                \left( \prod_{i=1}^{k-1}\superoperInjection_i \superoperMeasurement_i  \right) \, \rho_{0} .
    \end{equation}
Here and in the following, all super-operator products are ordered by decreasing indices. Note that
the expression \eqref{eq:rho_allKnowledge_ideal} can be computed iteratively from the recurrence
    \begin{eqnarray}\label{eq:filter_ideal}
        \rho_{k+1} =  \superoperMeasurement_{k+1} \superoperInjection_k  \rho_{k},
    \end{eqnarray}
setting $\superoperInjection_0 = \Id$. By applying this filter recursively at each cycle, we get a
reliable real-time estimate of the field state.

Based on Eqs.~\eqref{eq:Proj} and \eqref{eq:injection}, the super-operators are defined as
    \begin{eqnarray}\label{eq:superOperator_Proj_ideal}
        \superoperMeasurement_i \,\rho &=& \frac{  M_{s_i}  \rho  M_{s_i}^\dag }{\tr{M_{s_i} \rho M_{s_i}^\dag}}\,,\\
            \label{eq:superOperator_Displ_ideal}
        \superoperInjection_i\, \rho &=& D(\alpha_{i})\, \rho\, D(-\alpha_{i})\,.
    \end{eqnarray}
Here, $s_i=g$ or $e$ depending on the outcome of the $i^\mathrm{th}$ atomic measurement. The
control amplitude $\alpha_i$ is adjusted after each atom detection according to the feedback law
described in the next subsection.

We choose to prepare initially the cavity in a coherent state with a mean photon number equal to
the target value $\ngoal$. The phase of this field is used as a reference. As a consequence, its
amplitude $\alpha_0=\sqrt\ngoal$ is real. The phase $\phiR$ and the operator $\Phi(N)$, defining
the measurement super-operators $\superoperMeasurement_i$ through \eqref{eq:MgMe}, should be
adjusted to optimize the quantum filter~\eqref{eq:filter_ideal}. On the one hand, the mean
dephasing angle per photon $\bar{\phi}$ (estimated in the useful range of $n$ values) should be
large in order to resolve adjacent photon numbers. On the other hand, the measurement should lift
the ambiguity between a large range of possible photon numbers from $0$ to some $\nmax$, determined
by the spread of the initial coherent field. Due to the oscillating nature of the measurement
operators $M_s$, this sets a maximum value of $\bar{\phi}$ of the order of $\pi/\nmax$.

Finally, we need to optimally distinguish $\ngoal$ from $\ngoal\pm 1$. This is achieved by setting
the Ramsey phase $\phiR$  so that $(\phi_{\text{R}} + \Phi(\ngoal))/2 = \pi/4$. This setting
corresponds to a maximal variation of the atomic state detection probabilities around $\ngoal$. In
other words, we set the atomic Ramsey interferometer at mid-fringe when the cavity contains
$\ngoal$ photons.

\subsection{Feedback control}  \label{ssec:ideal:feedback}

The distance between $\rhohalf$ and the target Fock state described by a density operator
$\rhogoal=\ket{\ngoal}\bra{\ngoal}$ can be conveniently defined~as
    \begin{equation}\label{eq:Lyapounov}
        V(\rhohalf)=1-\tr{\rhohalf\rhogoal} \equiv 1 - F(\rhohalf),
    \end{equation}
where $F(\rho)=\tr{\rho\,\rhogoal}$ is the fidelity of a state $\rho$ with respect to the target
state $\rhogoal$. The function $V(\rho)$ represents a natural choice for the Lyapunov function
\cite{Mirrahimi07} used to study the stability properties of the quantum controller. This study is
presented in more detail in Ref.~\cite{Mirrahimi09}.

Let us denote the cavity state after the control displacement in the $k^\mathrm{th}$ iteration of
the feedback loop as $\rhoaux_k = \superoperInjection_k \rho_{k}$. The optimal convergence of the
feedback procedure is obtained by finding at each cycle the injection amplitude $\alpha_k$ which
maximizes the fidelity $F(\rhoaux_{k})$. This search can be performed in a straightforward way by
numerical optimization. However, this approach is time-consuming and could not be realistically
implemented with existing real-time data analysis systems in the few tens of microseconds time
interval between two atomic detections. We overcome this difficulty by developing a faster,
analytical calculation of a locally optimal displacement $\alpha_k$. Since the measurement
operators, the initial density matrix and the projector on the target Fock state are all real
operators, we can reasonably choose $\alpha_k$ to be real. The field density matrix then remains
always real.

In the limit of small $\alpha$, the Baker-Campbell-Hausdorff formula \cite{Barnett03} yields the
following approximation for a displaced state \eqref{eq:injection}:
    \begin{align}\label{eq:BKH}
        \tilde{\rho} = D(\alpha) \rho D(-\alpha) \approx \,& \rho -\alpha [\rho,a^\dag-a]\notag\\ & +
        \frac{\alpha^2}{2}[[\rho,a^\dag-a],a^\dag-a].
    \end{align}
Using~\eqref{eq:superOperator_Displ_ideal}, we therefore get for small $\alpha_k$
    \begin{multline}\label{eq:8}
        F(\rhoaux_{k}) = F(\rhohalf)
        - \alpha_k \tr{[\rhohalf,a^\dag\!-\!a]\rhogoal} \\
        + \frac{\alpha_k^2}{2}\tr{[[\rhohalf,a^\dag\!-\!a],a^\dag\!-\!a]\rhogoal}\,.
    \end{multline}
Choosing the feedback amplitude as
    \begin{equation}\label{eq:feed1}
        \alpha_k = - c_1 \tr{[\rhohalf,a^\dag\!-\!a]\rhogoal},
    \end{equation}
with a gain $c_1 >0$ small enough ensures that
    \begin{equation}
        F(\rhoaux_{k}) > F(\rhohalf)
    \end{equation}
as soon as the trace in \eqref{eq:feed1} is strictly positive. Furthermore, since $[\rhogoal,
M_g]=[\rhogoal, M_e]=0$ and $M_g^\dag M_g+M_e^\dag M_e=\openone$, the conditional expectation of
$F(\rho_{k+1})$ knowing $\rhoaux_{k}$ is given by
    \begin{multline}
        \EE{F(\rho_{k+1})~|~\rhoaux_{k}} = P_{g,k} \tr{\frac{\rhogoal M_g\rhoaux_{k} M_g^\dag}{P_{g,k}}}\\
        + P_{e,k} \tr{\frac{\rhogoal M_e\rhoaux_{k} M_e^\dag}{P_{e,k}}} = F(\rhoaux_{k}).
    \end{multline}
Therefore
    \begin{multline}
        \EE{ F\big(\rhoaux_{k+1}\big)~|~\rhoaux_{k}} \geq \EE{F\big(\rho_{k+1}\big)~|~\rhoaux_{k}} =F(\rhoaux_{k})
    \end{multline}
and, as a result, the expectation value of $F(\rhoaux_k)$ is a non-decreasing function of $k$:
    \begin{equation}
        1 \geq \EE{F\big(\rhoaux_{k+1}\big)} \geq \EE{F\big(\rhoaux_{k}\big)} \geq 0.
    \end{equation}

We have only shown so far that $F(\rhoaux_k)$ is increasing on the average. It does not mean that,
for {\sl all} individual realizations of the loop sequence, $F(\rhoaux_k)$ converges to 1, its
maximum value reached when $\rho$ is equal to $\rhogoal$. Indeed, the feedback law \eqref{eq:feed1}
does not prevent the convergence towards other Fock states, since $\alpha_k=0$ whenever $\rhohalf$
is the projector on any photon number state. A careful analysis shows that, in each realization,
$F(\rhoaux_k)$ converges either to 1 or, with some small probability, to zero. To overcome this
spurious attraction towards wrong photon number states, we propose, as in \cite{Mirrahimi07}, to
modify the feedback law~\eqref{eq:feed1} by applying a constant injection ``kick" as soon as the
cavity state significantly deviates from the target:
    \begin{equation}\label{eq:feedbackLaw}
        \alpha_k\!=\!\left\{\!
        \begin{array}{ll}
        c_1 \tr{[\rhogoal \, ,a^\dag\!-\!a]\rhohalf}     & \mbox{ if } F(\rho_{k}) \geq \varepsilon,\\
        \\
        c_2\, \text{sign}\left( \ngoal \!- n_{k}\right) & \mbox{ if } F(\rho_{k}) <    \varepsilon. \\
        \end{array}
        \right.
    \end{equation}
Here, $n_{k}$ is the mean photon number in the current state $\rhohalf$, $c_2>0$ is a constant kick
amplitude. The kick condition is set by $\varepsilon$ with $1\! \gg\! \varepsilon\!
>\! 0$. In this way, by breaking attraction towards other Fock states, we make sure that  the
cavity state converges towards the target one.

The controller gain $c_1$ in \eqref{eq:feedbackLaw} must be tuned to maximize fidelity
$F(\rhoaux_{k})$ at each sampling time $k$. Up to third-order terms in $(\rhohalf\!-\rhogoal)$,
Eq.~\eqref{eq:8} yields
    \begin{multline}
        F(\rhoaux_{k})  = F(\rhohalf)\, +\\
        \Big(\tr{[\rhogoal,a^\dag\!-\!a]\rhohalf}\!\Big)^2 \left( c_1\! -\! \frac{c_1^2}{2}
            \tr{[\rhogoal,a^\dag\!-\!a]^2}\!\right)\!.
    \end{multline}
By maximizing this expression, we find that the gain
    \begin{equation}\label{eq:c1}
        c_1 = \tr{[\rhogoal ,a^\dag\!-\!a]^2}^{-1}\! = (4\ngoal\!+2)^{-1}
    \end{equation}
provides the fastest convergence speed in the vicinity of $\rhogoal$. A detailed mathematical analysis and a
convergence proof of this feedback-scheme will be given in \cite{Mirrahimi09}.

\subsection{Simulations} \label{ssec:ideal:simu}

    \begin{figure}[!t]
        \centerline{ \includegraphics[width=0.47\textwidth]{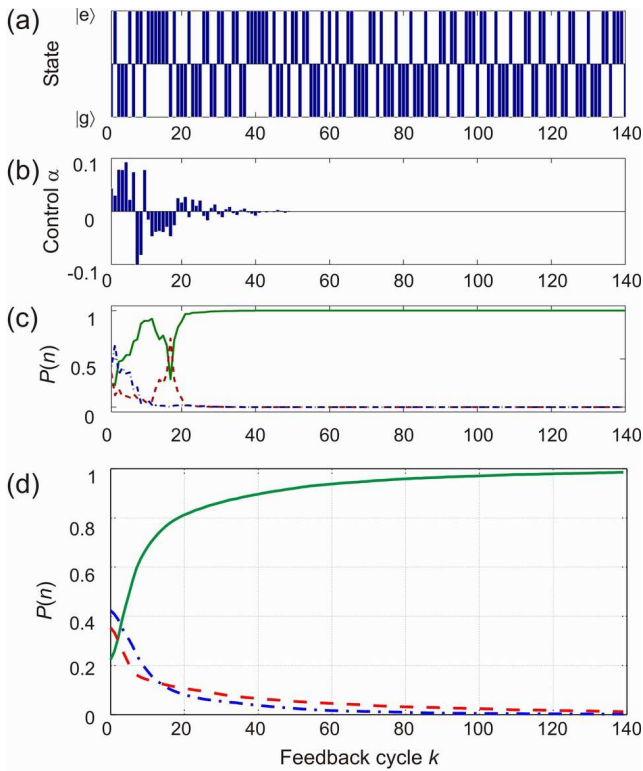}}
        \caption{A single closed-loop quantum trajectory of an ideal measurement
        with~$|\ngoal\rangle=|3\rangle$. (a) Atomic states detected at each feedback cycle. (b)
        Control injections $\alpha_k$. (c) Evolution of photon number probabilities
        $P(n\!<\!\ngoal)$, $P(\ngoal)$ and $P(n\!>\!\ngoal)$ shown as dashed-dot blue, solid green and dashed red curve,
         respectively. (d) Average over $10^4$ closed-loop quantum trajectories.}
        \label{fig:TrajIdeal}
        \label{fig:MeanIdeal}
    \end{figure}

To assess the performance of the proposed feedback scheme, we have performed quantum Monte-Carlo
numerical simulations. Figure~\ref{fig:TrajIdeal}(a)-(c) presents a single quantum closed-loop
trajectory in the ideal measurement setting, with a target Fock state $|\ngoal\rangle=|3\rangle$.
The Hilbert space is limited to $\nmax=9$ photons. The atom-cavity detuning is set to $\delta/2\pi
= 238$~kHz. The mean dephasing angle per photon around $\ngoal$ is then about $\pi/7$. Intuitively,
the Ramsey phase $\phiR$ should be set at the mid-fringe setting for $\ngoal$ photons, i.e.,
$(\phi_{\text{R},0} + \Phi(\ngoal))/2 = \pi/4$. We have numerically observed that the convergence
of the process is faster when the Ramsey phase alternates between four values in successive
feedback cycles: $\phi_{\text{R},0}$, $(\phi_{\text{R},0}+\sigma_\text{R})$, $\phi_{\text{R},0}$
and $(\phi_{\text{R},0}-\sigma_\text{R})$. The phase excursion $\sigma_\text{R}$ is set at
$0.69$~rad. The kick amplitude $c_2$ is  0.1 with a kick zone defined by $\varepsilon=0.1$. The
initial cavity field $\rho_0$ is the coherent state with 3 photons on the average,
$\rho_0=D\left(\sqrt{3}\right) \ket{0}\bra{0} D\left(-\sqrt{3}\right)$.

The upper trace in Fig.~\ref{fig:TrajIdeal} shows a sequence of detected atomic state. The
corresponding control inputs $\alpha_k$ are plotted in the next panel. Large at the beginning,
$\alpha_k$ rapidly converges towards zero. Figure~\ref{fig:TrajIdeal}(c) shows the probability for
the number of photons in the cavity field to be smaller, equal or larger than $\ngoal$ (dashed-dot
blue, solid green and dashed red curve, respectively). For this particular quantum trajectory, we
observe that $\rho$ converges to $\rhogoal$ in less than $30$ feedback cycles.

Figure~\ref{fig:MeanIdeal}(d) presents an ensemble average over $10^4$ realizations of the same
numerical feedback experiment. The fidelity $F(\rhoaux_k)$ reaches values above $0.99$ after $140$
cycles. After 20 cycles only, about $80\%$ of the trajectories have converged to the target state,
illustrating the efficiency of the feedback method.

\section{Realistic experiment} \label{sec:real}

So far, we have neglected experimental imperfections. In this Section, we take into account all
known imperfections of the present experimental set-up: finite cavity lifetime, Poisson
distribution of the atom number in atomic samples, non-ideal efficiency and state-selectivity of
the detector and, finally, the finite delay between atom-cavity interaction and atomic detection.
Below, we introduce modifications of the quantum filter~\eqref{eq:filter_ideal}, which make the
process efficient in spite of these imperfections.

\subsection{Modified quantum filter}

The quantum filtering process must provide us with a density matrix describing all our knowledge of
the cavity field at a given time. We chose here this time to be right before the injection of the
actuator pulse in the $k^\mathrm{th}$ iteration of the loop, after detection of the $k^\mathrm{th}$
atom. At this time, our knowledge includes all atomic measurements so far (from 1 to $k$), all
displacements performed and the known relaxation of the cavity field in between these events. It
must also include the influence of the $d$ atomic samples which have interacted with the cavity,
but are still flying towards the detector when we record atomic sample $k$. All this information
can, as above, be represented by super-operators acting on the field's density matrix.

Let us note $\superoperProjection_s$ the super-operator describing the detection of an atomic
sample. Each detection has three possible outcomes, labelled $s=e$, $g$ or $u$ (atom detected in
$e$, $g$ or no detected atom at all). The super-operator $\superoperNoprojection$  describes the
action of one of the $d$ atomic samples flying from $C$ to $D$. The displacement super-operator
$\superoperInjection$ is given by \eqref{eq:injection} and the field relaxation during the time
interval between atomic samples is described by $\superoperEvolution$. The exact expressions of all
these super-operators will be given later.

Therefore, the state of the cavity field before injection~is
    \begin{equation}\label{eq:rho_allKnowledge}
    \rho_{k} = \superoperUndetected{,k} \, \superoperDetected{,k} \, \rho_{0},
    \end{equation}
where
    \begin{equation}\label{eq:superoperatorF1}
        \superoperDetected{,k}   = \superoperEvolution \superoperProjection_{s_k} \prod_{i=1}^{k-1} \superoperInjection_{i-d} \superoperEvolution \superoperProjection_{s_i}
    \end{equation}
includes the information gathered by the first $k$ sample detections and
    \begin{equation}\label{eq:superoperatorF2}
        \superoperUndetected{,k} = \prod_{i=k}^{k+d-1} \superoperEvolution \superoperNoprojection \superoperInjection_{i-d}
    \end{equation}
includes the influence of the $d$ in-flight samples. Note that due to the atomic propagation delay
between $C$ and $D$, the displacement amplitude injected in $C$ right after atomic sample $i$ has
interacted with it is $\alpha_{i-d}$, computed in the $(i-d)$-th iteration of the loop. Hence, the
first $d$ atomic samples cross the cavity before any displacement. This is formally taken into
account by setting $\superoperInjection_i$ to unity for non-positive indices. If $d=0$, i.e.~for no
delay in the detection process, the empty product in \eqref{eq:superoperatorF2} equals by
convention the identity operator. Consequently, in the case of the ideal experimental parameters,
the expression \eqref{eq:rho_allKnowledge} naturally reduces to \eqref{eq:rho_allKnowledge_ideal}.

The density operator $\rho_k$ is used in the feedback law \eqref{eq:feedbackLaw} to calculate the
control injection $\alpha_{k}$. The product $\superoperDetected{,k}$ can be easily computed
recursively, as in the ideal case. However, we must recalculate $\superoperUndetected{,k}$ at each
feedback cycle.

We now give the explicit expression of all super-operators considered so far. The measurement
process $\superoperProjection_s$ takes into account detector's imperfections and non-ideal atomic
source. If the detector clicks and indicates state $|s\rangle=|g\rangle$ or $|e\rangle$, the new
quantum state is a statistical weighted mixture of two projected states. The strongest weight
corresponds to the projection according to the recorded outcome. With a smaller probability, the
atomic detection has been erroneous. The atom was, at detection time, in the state $\ket
{\overline{s}}$, opposite to $\ket s$, and the field has been projected accordingly. Thus, due to
the imperfections of state discrimination alone, we get
    \begin{equation}\label{eq:rho_click}
    \superoperProjection_{(s=e/g)} \,\rho_{} = (1-P_\text{f})\superoperMeasurement_s \rho
                    + P_\text{f} \superoperMeasurement_{\overline{s}} \rho \, .
    \end{equation}
The weights of the two projected states are given by the conditional probability of a wrong
detection knowing that the detector has clicked in $|s\rangle$: $P_\text{f}= \etaf
P_{\overline{s}}/\left((1-\etaf) P_{s}+ \etaf P_{\overline{s}}\right)$ with $P_s=\tr{M_s \rho
M_s^\dag}$.

If the detector does not click (outcome $s=u$), either there was no atom in the sample or an atom
was present but has not been detected. Therefore, the estimated state is a mixture of the
unperturbed cavity field and of the two projected states corresponding to the two possible states
of the undetected atom (we assume here that these two states are equiprobable). We thus get
    \begin{equation}\label{eq:rho_noclick}
        \superoperProjection_{(s=u)}\, \rho = (1\!-\!P_{u}) \rho +
        P_{u} \left(M_g\rho M_g^\dag\! + \!M_e\rho  M_e^\dag \right).
    \end{equation}
The weights in this mixture are set by the conditional probability to have an undetected atom in a
sample:  $P_{u}= \etap(1-\etad)/(1-\etap\etad)$.

The interaction with a not-yet-detected sample is described by the super-operator
$\superoperNoprojection$. It is given by  \eqref{eq:rho_noclick} with the conditional probability
$P_{u}$ replaced by the atomic sample occupation probability $\etap$, since the measurement has not
been performed yet, resulting in
    \begin{equation}\label{eq:rho_nodetection}
        \superoperNoprojection \rho = (1\!-\!\etap) \rho +
        \etap \left(M_g\rho M_g^\dag\! + \!M_e\rho  M_e^\dag \right).
    \end{equation}

The super-operator $\superoperEvolution$ describes the evolution due to the cavity field relaxation
during the time interval $\Tp$ between two atomic samples. Using \eqref{eq:L} in the approximation
of small time interval, $\Tp\! \ll\! \Tcav$, we get
    \begin{equation}\label{eq:rho_jump}
        \superoperEvolution \rho = (\openone + \Tp \textbf{L}) \rho \,.
    \end{equation}

\subsection{Simulations}

    \begin{figure}[!t]
        \centerline{\includegraphics[width=0.45\textwidth]{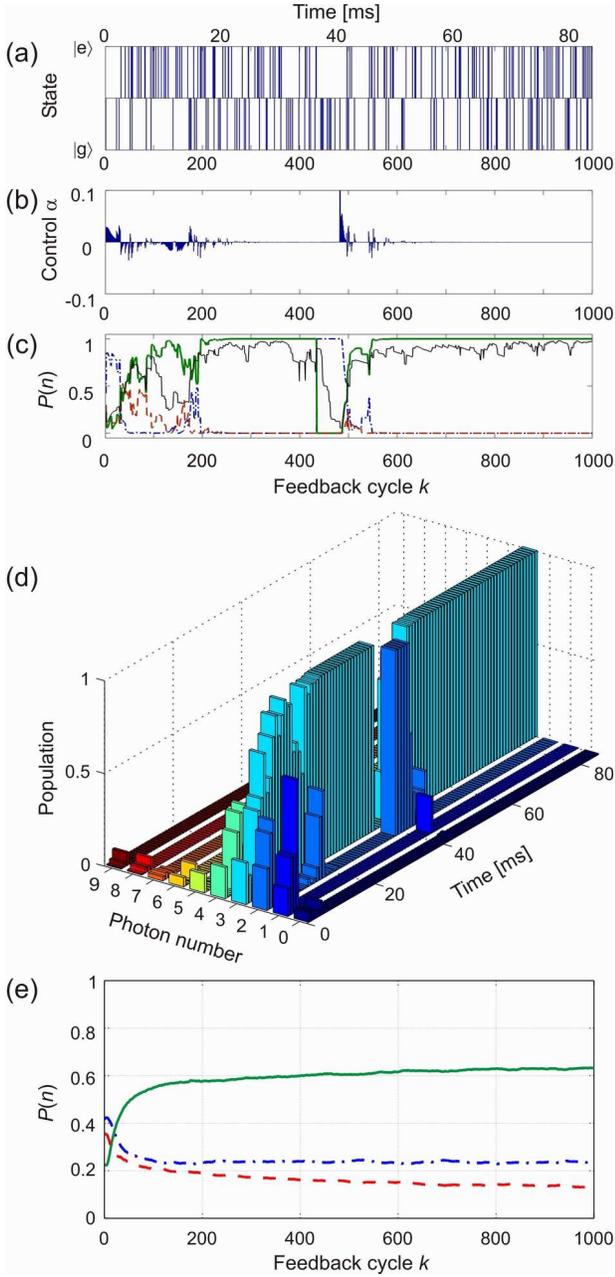}}
        \caption{A single closed-loop quantum trajectory for realistic experimental parameters
        with~$|\ngoal\rangle=|3\rangle$. (a) Sequence of detected atomic states. (b) Control injections
        $\alpha_k$. (c) Evolution of the photon number probabilities $P(n\!<\!\ngoal)$, $P(\ngoal)$ and
        $P(n\!>\!\ngoal)$ (dashed-dot blue, solid green and dashed red curve, respectively) of the field
        state $\rhoreal$ deduced from the Monte-Carlo simulation. The thin black line gives the estimated state
        fidelity $F(\rho_k)$. (d) Photon number probabilities for $n=0$ to $9$ in $\rhoreal$. For
        better clarity, results of only one out of each 10 feedback cycles are shown. (e) Average over
        $10^4$ closed-loop quantum trajectories. If the number of quantum trajectories is large enough
        (as is the case here), the averaged photon number probabilities calculated from the real field
        states $\rhoreal$ coincide with those averaged over the estimated states. }
        \label{fig:TrajReal}
        \label{fig:MeanReal}
    \end{figure}

We have performed extensive simulations, similar to those of the idealized measurement in
Fig.~\ref{fig:TrajIdeal}, with the feedback algorithm adapted to the realistic experimental
parameters, given in Sec.~II. We use Monte-Carlo simulations to follow the state of the cavity
along individual realizations of the experiment. The field relaxation is taken into account by
simulating random quantum jumps, whose average effect is described by Eq.~\eqref{eq:rho_jump}
\cite{Haroche06}. For each atomic sample, we choose randomly the number of atoms ($0$ or $1$) and
the real atomic state ($\ket e$ or $\ket g$) which would be detected by the ideal detector. These
results determine uniquely the subsequent field state $\rhoreal_k$. We then take into account the
detection imperfections to simulate the outcomes of the realistic measurement, which are used in
the feedback loop for computing $\rho_k$. Note that $\rhoreal$ is not accessible in the experiment.
Here, we use it to simulate the evolution of the field and, by comparing it with the state
estimated in the quantum filtering process, to characterize the feedback performance. In the ideal
case, considered in Sec.~\ref{sec:ideal}, the estimated state $\rho_k$ always coincide with the
real state of the field $\rhoreal_k$.

Figure~\ref{fig:TrajReal}(a)-(d) shows a typical quantum trajectory, illustrating the robustness of
our feedback scheme at the present level of imperfections. The first difference with the ideal case
is the presence of many feedback cycles with no detected atoms. Nevertheless, after about
$k\!=\!100$ feedback cycles (each of $85\,\mu$s duration), corresponding to approximatively
$k\etap\!\approx\! 30$ atoms and $k\etap\etad \!\approx\! 24$ detector clicks, the cavity state
successfully converges to the target Fock state ~$|\ngoal\rangle = |3\rangle$.

The other striking feature is the presence of sudden photon jumps due to the limited cavity
lifetime. In this particular trajectory, a photon loss occurred at feedback cycle number
$k\!\approx\!430$ (at about 37~ms). One sees that about 60 cycles (about 5~ms) are required to
detect the jump before applying a large control displacement and about 60 more cycles to restore
the target photon number in the cavity field.

    \begin{figure}[!t]
        \centerline{\includegraphics[width=0.48\textwidth]{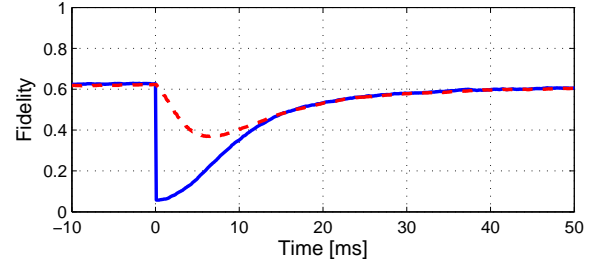}}
        \caption{Feedback performance in the presence of a sudden photon number loss.
        The fidelity of the actual cavity state $\rhoreal_k$ (blue solid curve) and of the state estimated by the quantum
        filtering process $\rho_k$ (red dashed curve) are defined
        with respect to the target Fock state and are obtained by averaging over $10^4$ quantum trajectories.
        In each trajectory, the time origin is shifted to the time of a quantum jump.}
        \label{fig:FidelityAfterJump}
    \end{figure}

Figure~\ref{fig:MeanReal}(e) shows the same type of ensemble average as presented in
Fig.~\ref{fig:MeanIdeal}(d). We observe a convergence towards a fidelity of $0.63$ after several
hundred cycles resulting in about 100 detector clicks. This convergence, given as the number of
actually detected atoms, is similar to that observed in the ideal case. The main difference between
the two traces is the reduced asymptotic value of the fidelity of the prepared state: $F = 0.63$
instead of $F = 1$ in the ideal case. This reduction is mainly due to the cavity decay, which
results in uncontrolled jumps of the photon number. The observed fidelity indicates that for a
typical closed-loop trajectory the field stays in the state $|3\rangle$ about $63\,\%$ of time.
Since the lifetime of $|\ngoal\rangle$ is given by $\Tcav/((1+\nth)\ngoal + \nth(n+1)) = 45$~ms
\cite{Brune08}, we estimate that it takes on the average 26~ms to restore the initial Fock state
after a sudden jump.

In order to observe in more detail the recovery of the system after a sudden photon jump, we
recorded $10^4$ quantum trajectories showing sudden photon losses at different times. Before
averaging the trajectories, we shift their individual time origins to the time of a jump.
Figure~\ref{fig:FidelityAfterJump} shows the evolution of the fidelity of the actual (solid curve)
and estimated (dashed curve) cavity states with respect to $\rhogoal$. As calculated before, it
takes about $20 - 30$~ms to restore the initial photon number state after a sudden photon loss.

\begin{figure}
        \centerline{\includegraphics[width=0.48\textwidth]{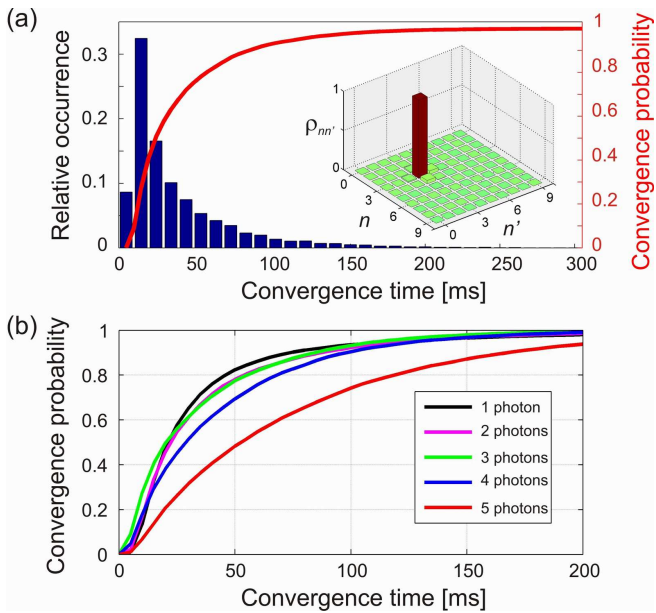}}
        \caption{Feedback convergence. (a) Histogram of convergence times with $\Fconv = 0.95$.
        The solid line gives the probability of the feedback convergence after a given time.
        The inset shows the average density matrix of the converged states.
        (b) Convergence for several different Fock states.
        These curves result from an average over $10^4$ quantum trajectories.}
        \label{fig:Success}
    \end{figure}

The maximum fidelity of the quantum feedback, seen in
Figs.~\ref{fig:MeanReal}-\ref{fig:FidelityAfterJump} and corresponding to the ensemble of many
realizations, shows how well a chosen photon number state can be preserved on average from its
unavoidable decay. However, since in each feedback cycle we acquire actual information on the
field, the fidelity of the deterministic generation of Fock states in a single realization can be
much higher. To reliably produce the target state $\rhogoal$ we keep the feedback process running
until we detect its successful convergence to $\rhogoal$ with an estimated fidelity better than
$\Fconv$. Figure \ref{fig:Success}(a) shows a histogram of convergence times for $\Fconv = 95\,\%$
obtained from $10^4$ quantum trajectories. The cumulative distribution, presented by a solid line,
shows that after 20~ms the probability of a successful feedback outcome reaches $50\,\%$ and that
it exceeds $90\,\%$ after 85~ms of the feedback operation. The inset in Fig.~\ref{fig:Success}(a)
shows the average density matrix of converged states with the population in $|\ngoal\rangle$ of
about $95\,\%$.

Our feedback scheme can be tuned to generate any small photon number states. Figure
\ref{fig:Success}(b) illustrates the convergence for several Fock states from $|\ngoal\rangle =
|1\rangle$ to $|5\rangle$. By increasing $\ngoal$, the initial photon number distribution gets
broader as the photon number lifetime gets shorter. Therefore, the feedback performance slowly
degrades, resulting in the increased convergence time for the same required fidelity~$\Fconv$.

\subsection{Real-time computation}

Implementation of this quantum feedback scheme into a real experiment requires fast real-time
analysis of the measured data, faster than duration $\Tp$ of one feedback cycle. We have tested a
real-time processing system $\mathrm{ADwin}$ (J\"{a}ger Messtechnik) allowing for fast data
acquisition and analysis in combination with complex experiment control. After careful optimization
of all feedback computations, the execution of one typical feedback cycle of the proposed scheme
requires about 4000 floating-point operations. Our preliminary tests have revealed that the ADwin
system needs for this task about $30\,\mu\mathrm{s} < \Tp = 85\,\mu$s. We are now working on the
integration of this control system into our cavity QED set-up.

\section{Conclusion}\label{sec:conclusion}

We have presented a quantum feedback protocol designed to deterministically generate small photon
number states of a trapped microwave field. The reliable estimation of the cavity state at each
feedback cycle allows us to follow in real time the quantum jumps of the photon number and then to
efficiently compensate them, thus protecting Fock states against decoherence.

The two main components of the feedback algorithm are the quantum filter, which estimates the
actual state of the field based on the outcomes of QND measurements, and the feedback law computing
the amplitude of a correction microwave pulse whose injection into the cavity mode maximizes the
field fidelity with respect to a desired Fock state. Quantum Monte-Carlo simulations of the QND
measurements and the quantum feedback response demonstrate the high reliability of our closed-loop
scheme even in the presence of realistic experimental imperfections. This scheme is also robust
with respect to an imprecise knowledge of the initial state, since the repeated measurements of the
state corrects initial wrong or incomplete information. Convergence and stability proofs of these
feedback schemes will be given elsewhere. For a first set of mathematical results, see
Ref.~\cite{Mirrahimi09}.

The presented scheme can be extended to include the Ramsey phase $\phiR$ and the spin dephasing
$\Phi(N)$ as additional control parameters of the feedback loop. This should help to optimize the
QND measurement according to our estimate of the actual field  state and thus
to acquire more information on its dynamics. These studies are under the way.\\

{\bf Acknowledgements} Work supported partially by Agence Nationale de la Recherche (projet
ANR-05-BLAN-0200-01 and projet Blanc CQUID 06-3-13957), Japan Science and Technology Agency (JST)
and EU (IP project SCALA).


\begin{thebibliography}{99}

\bibitem{Guerlin07}
C.~Guerlin {\it et al.},
Nature {\bf 448}, 889 (2007)

\bibitem{Wiseman94}
H.~M.~Wiseman and G.~J.~Milburn,
Phys.~Rev.~A~{\bf 49}, 4110 (1994)

\bibitem{Deleglise08}
S.~Del\'{e}glise {\it et al.},
Nature {\bf 455}, 510 (2008)

\bibitem{Belavkin92}
V.~P.~Belavkin,
J.~Multivariate~Anal.~{\bf 42}, 171 (1992)

\bibitem{Bouten07}
L.~Bouten, R.~Van Handel and M.~James,
SIAM J.~Contr.~Optim.~{\bf 46}, 2199 (2007)

\bibitem{Mirrahimi07}
M.~Mirrahimi, R.~Van Handel,
SIAM~J.~Contr.~Optim.~{\bf 46}, 445 (2007)

\bibitem{Geremia06}
J.~M.~Geremia,
Phys.~Rev.~Lett.~{\bf 97}, 073601 (2006)

\bibitem{Mirrahimi09}
M.~Mirrahimi, I.~Dotsenko and P.~Rouchon,
in preparation (quant-ph/0903.0996)

\bibitem{Raimond01} J.-M.~Raimond, M.~Brune, and S.~Haroche,
Rev.~Mod.~Phys.~{\bf 73}, 565 (2001)

\bibitem{Brune08}
M.~Brune \textit{et al.},
Phys.~Rev.~Lett.~{\bf 101}, 240402 (2008)

\bibitem{Walls94}
D.~F.~Walls and G.~J.~Milburn, {\em  Quantum Optics\/}, (Springer, Berlin, 1994).

\bibitem{Barnett03}
S.~M.~Barnett and P.~M.~Radmore, {\em  Methods in Theoretical Quantum Optics\/},
    (Oxford University Press, 2003).

\bibitem{Haroche06}
S.~Haroche and J.-M.~Raimond, {\em  Exploring the Quantum\/},
(Oxford University Press, 2006).

\end{thebibliography}
\end{document}